\documentclass[12pt,a4paper,reqno]{amsart}
\usepackage{xypic,a4wide,amsmath,amssymb,amsthm}
\usepackage[utf8x]{inputenc}

\usepackage[T1]{fontenc}

\usepackage{amssymb}
\usepackage{amsmath}

\usepackage[dvips]{graphicx}
\usepackage{array}

\title[\ ]{Quasideterminant solutions of NC 
Painlev\'e II equation with the Toda solution at $ n=1 $ as a seed solution in its Darboux transformation}

\author{Irfan Mahmood}
\address{
 (1)-Department of Mathematics, Universit\'e d'Angers,\\
2 Boulevard Lavoisier, 49045 Angers Cedex 01, France\\
(2)- University of the Punjab, 54590 Lahore,  Pakistan} \email{mahirfan@yahoo.com} \urladdr{}

\begin{document}

\maketitle

\begin{abstract}
In this  paper, I construct the Darboux transformations for the non-commutative Toda solutions  at $ n=1 $ with the help of  linear systems whose compatibility condition yields zero curvature representation of  associated systems of  non-linear differential equations.  I also  derive the  quasideterminant solutions of the non-commutative Painlev\'e II  equation by taking the  Toda solutions at $ n=1 $ as a seed solution in its Darboux transformations. Further by iteration, I generalize the Darboux transformations of the seed solutions to $ N$-th form. At the end I describe the zero curvature representation of quantum Painlev\'e II equation that involves Planck constant  $ \hbar  $ explicitly and system reduces to the classical Painlev\'e II  when $ \hbar \rightarrow 0  $.
\end{abstract}
\section{Introduction}

The Painlev\'e equations were discovered by Painlev\'e and his colleagues when they have  classified  the  nonlinear second-order ordinary differential equations with respect to their solutions \cite{r1}.The study of Painlev\'e equations is important from  mathematical point of view  because of their frequent appearance in the  various areas of physical sciences including plasma physics, fiber optics, quantum gravity and field theory, statistical mechanics, general relativity and nonlinear optics. The classical Painlev\'e equations are regarded as completely integrable equations and obeyed the Painlev\'e test \cite{r5, r11, r12}.
These equations admit  some properties such as linear representations, hierarchies, they possess Darboux transformations(DTs)  and Hamiltonian structure. 
These equations  also arise as ordinay differential equations (ODEs)   reduction of some integrable systems, i.e, the ODE reduction of the KdV equation is Painlev\'e II (PII) equation \cite{r4, MA}.\\

The noncommutative(NC) and quantum extension of Painlev\'e equations  is  quite interesting  in order to explore the properties  which they  possess 
 with respect to usual Painlev\'e systems on ordinary spaces.
NC spaces are characterized by the noncommutativity of the spatial co-ordinates. For example, if $ x^{\mu} $ are the space co-ordinates then the noncommutativity is defined by
 $ [x^{\mu},x^{\nu}]_{\star}=i\theta^{\mu \nu} $
where parameter $ \theta^{\mu\nu}$ is anti-symmetric tensor and Lorentz invariant and $ [x^{\mu},x^{\nu}]_{\star}$ is commutator under the star product. NC field theories on flat spaces are given by the replacement of ordinary
products with the Moyal-products and realized as
deformed theories from the commutative ones. Moyal product for ordinary fields $ f(x)$  and $ g (x)$ is  explicitly defined by
\[f(x) \star g(x)=exp(\frac{i}{2}\theta^{\mu\nu} \frac{\partial}{\partial x^{'\mu}}\frac{\partial}{\partial x^{''\nu}})f(x^{'})g(x^{''})_{x=x^{'}=x^{''}}\]
\[=f(x)g(x)+\frac{i}{2}\theta^{\mu\nu}\frac{\partial f}{\partial x^{'\mu}}\frac{\partial g}{\partial x^{''\nu}}+ \mathcal{O}(\theta^2).\]this product obeys associative property
$f\star ( g \star h)= (f \star g) \star h$,
if we apply the commutative limit
$\theta^{\mu \nu} \rightarrow 0 $
then above expression will reduce to ordinary product as   $ f \star g = f. g.$ In our case the NC product is the Moyal-product and we consider the non-commutativity between  space variable and its function. \\

We are familiar with Lax equations as a nice representation of integrable systems. The  Lax equation and zero curvature condition both have same form on  deformed spaces  as they possess on ordinary space. These representations involve two linear operators, these operators may be differential operators or matrices \cite{r13}- \cite{r24}.
If $A$ and $B$ are the linear operators then Lax equation is given by  $ A_{t} = [B,A]$  where $[B,A]$
is commutator under the star product or quantum product, this Lax pair formalism is also helpful to construct the DT, Riccati equation and BT  of integrable systems. The compatibility condition of inverse scattering problem  $\Psi_{x}=A(x,t)\Psi$ and $\Psi_{t}=B(x,t)\Psi$ yields $ A_{t}-B_{x}= [B,A]$  which is called the zero curvature representation of integrable systems \cite{e2}-\cite{4}. Further we will denote the commutator  and anti-commutator by $[,]_{-}$ and  $[,]_{+}$ respectively. Now the Lax equation and zero curvature condition can be expressed as  $ A_{t} = [B,A]_{-}$ and  $ A_{t}-B_{x}= [B,A]_{-}$. \\

The  Painlev\'e equations can be represented by the Noumi-Yamada systems  \cite{NY}, these systems are discovered by Noumi and Yamada
while studying  symmetry of Painlev\'e equations and these systems also possess the affine  Weyl group symmetry of type $ A^{1}_{l}$. For example Noumi-Yamada system
for Painlev\'e II equation is given by

\begin{equation}\label{eq:1} 
\left\{
\begin{array}{lr}
  u^{'}_{0}= u_{0}u_{2} + u_{2}u_{0} + \alpha_{0} \\
  u^{'}_{1}= -u_{1}u_{2} - u_{2}u_{1} + \alpha_{1}\\
u^{'}_{2}= u_{1} - u_{0}
\end{array}
\right.
 \end{equation} 
where $ u^{'}_{i} = \frac{d u_{i}}{dz}$  and $ \alpha_{0}$ , $ \alpha_{1}$  are constant parameters. Above system \ref{eq:1} also  also a unique representation of NC and quantum PII equation, for NC derivation of PII equation \cite{7} the dependent functions $ u_{0} $, $ u_{1} $,  $ u_{2} $  obey a kind of star product and in case of quantum derivation these functions are subjected to some quantum commutation relations \cite{NH} and \cite{NGR}.\\ 
 In this  paper, I construct the Darboux transformations for the solutions of Toda equations at $ n=1 $, $ u_{1} = \phi^{'} \phi^{-1}$ and its negative counterpart $ u_{-1} = \psi^{'} \psi^{-1}$, with the help of linear systems whose compatibility condition yields zero curvature representation of their associated systems of  non-linear differential equations.  I also  derive the  quasideterminant solutions of the non-commutative Painlev\'e II  equation by taking the  Toda solutions at $ n=1 $ as a seed solution in its Darboux transformations. Further by iteration I generalize the Darboux transformations of the seed solutions  of the NC PII equation to the $ N$-th form. 
I also describe an equivalent zero-curvature representation of quantum PII equation that involves Planck constant  $ \hbar  $ explicitly. Further, I construct the quantum PII Riccati form with the help of its linear system by using the method of Konno and Wadati \cite{KKMW}.

\section{Brief introduction of Non-commutative Painlev\'e II equation}
The following NC analogue of classical Painlev\'e II equation 
\begin{equation}\label{NCPIIa} 
  u_{2}^{''} = 2u^{3} - 2 [ z, u ]_{+}  + C
\end{equation} 
where  $ [ z, u ]_{+}  = zu  + u  z $ and constant $ C= 4 ( \beta + \frac{1}{2}) $ was obtained by eliminating $ u_{0} $ and $ u_{1} $ from (\ref{eq:1}), here $ u = u_{2}$  \cite{7}.
Further it was shown by  V. Retakh and V. Roubtsov  that with the following identities 
 \begin{equation}\label{TODA2} 
\phi^{''} \phi^{-1}= 2 z-2 \phi \psi 
\end{equation}
\begin{equation}\label{TODA3} 
\psi^{-1} \psi^{''} = 2z - 2  \phi \psi 
\end{equation}
and 
\begin{equation}\label{TODA4} 
\psi \phi^{'} - \psi^{'} \phi= 2\beta
\end{equation}
the solutions $ u_{n}  = \theta_{n}^{\prime} \theta_{n}^{-1} $  of the Toda equation 
\begin{equation}\label{TODA1} 
(\theta_{n}^{\prime} \theta_{n}^{-1} )^{\prime}=  \theta_{n+1} \theta_{n}^{-1} - \theta_{n} \theta^{-1}_{n-1} \;  \text{for} \; n \geqslant 1 
\end{equation} 
 satisfies  the NC PII$(z, \beta +n- 1 ) $ equation and  the  solutions  $ u_{-m}  = \eta_{-m}^{\prime} \eta_{-m}^{-1} $  of the negative counter part of  (\ref{TODA1}) 
\begin{equation}\label{TODA1n} 
(\eta^{-1}_{-m} \eta^{\prime}_{-m})^{\prime}=  \eta_{-m}^{-1} \eta_{-m-1}-  \eta_{-m}+1^{-1} \eta_{-m}  \;   \text{for}  \;  m \geqslant 1
\end{equation} 
 satisfies  the  NC PII $ (z, \beta -n )$  equation, here $ \theta_{1} = \phi ,  \theta_{0} = \psi^{-1}  $ and $ \eta_{0} = \phi^{-1} , \eta_{-1} = \psi  $.  In the following section we review the zero curvature representation of NC PII equation   (\ref{NCPIIa}). Further in proposition 1.1, we construct the linear representation of   (\ref{TODA2})  and   (\ref{TODA3})  that will be helpfull to derive an explicit expression of the  Darboux transformations for $ \phi $ and $ \psi$.. 
\subsection{Zero curvature representation of NC PII equation}
The NC PII equation (\ref{NCPIIa}) can  be derived from inverse scattering problems with zero constant $C=0$ \cite{BC} and  in general form with non zero constant $ C \neq 0 $ \cite{MIRFAN}.
Let consider the following linear system \cite{MIRFAN} 
\begin{equation}\label{NCPIIb} 
 \Psi_{\lambda}=A(z;\lambda)\Psi \; \;  \Psi_{z}=B(z;\lambda)\Psi 
\end{equation} 
with Lax matrices
\begin{equation}\label{RDTa}
\left\{
\begin{array}{lr}
A= (8i \lambda^{2} + iu^{2} - 2iz ) \sigma_{3} + u^{'} \sigma_{2}+  (\frac{1}{4} C\lambda^{-1} -4\lambda u )\sigma_{1} \\
B =  -2i \lambda \sigma_{3}  + u \sigma_{1}  
\end{array}
\right.
 \end{equation} 
where $ \sigma_{1}$, $ \sigma_{2}$ and $ \sigma_{3}$ are Pauli spin matrices given by
 \[   \sigma_{1} = \begin{pmatrix}
0 & 1 \\
1 & 0
\end{pmatrix}, \; \sigma_{2} =\begin{pmatrix}
0 & -i \\
i & 0
\end{pmatrix} \; \sigma_{3}  = \begin{pmatrix}
1 & 0 \\
 0 & -1
\end{pmatrix}\]
here $ \lambda $ is spectral parameter. The linear system (\ref{NCPIIb}) is a equivalent representation of NC PII equation the compatibility condition of that system yields NC PII equation (\ref{NCPIIa}).
The following proposition contains the derivation of NC PII Riccati form by using the method of Konno and Wadat  \cite{KKMW}\\
\textbf{Proposition 1.1.}\\
The linear system (\ref{NCPIIb}) with eigenvector $ \Psi = \begin{pmatrix}
\chi  \\
\Phi
\end{pmatrix}$ and setting $ \Gamma = \chi \Phi^{-1}$ can be reduced to the following NC PII Riccati form
 \[ \Gamma^{'} = -4i\lambda \Gamma +u - \Gamma u \Gamma \]
\textbf{Proof:}\\
In order to derive the NC PII Riccati we consider following  the eigenvector 
$ \Psi = \begin{pmatrix}
\chi  \\
\Phi 
\end{pmatrix}$ in linear systems (\ref{NCPIIb}) and  we obtain

\begin{equation}\label{RENCPII1}  
 \left \{ \begin{aligned}
\frac{d \chi}{d \lambda} = ( 8  i \lambda^{2} +  i u^{2}-2i z) \chi + ( -i u_{z} + \frac{1}{4} C \lambda^{-1} - 4 \lambda u ) \Phi \\
\frac{d \Phi}{d \lambda} = (i u_{z}+\frac{1}{4}C \lambda^{-1}-4\lambda u )\chi + (-8  i\lambda^{2} - i u^{2}+2i z)\Phi
\end{aligned}
 \right. 
 \end{equation}
and 
\begin{equation}\label{RENCPII2}  
\left \{ \begin{aligned}
\chi^{'} = -2i \lambda  \chi + u \Phi \\
\Phi^{'} =u \chi + 2i \lambda \Phi
\end{aligned}
 \right. 
\end{equation}
where $ \chi^{'} = \frac{d \chi}{dz}$ and  from system (\ref{RENCPII2}) we can evaluate the following expressions 
 \begin{equation}\label{RENCPII3} 
  \chi^{'} \Phi^{-1} =  -2i \lambda  \chi \Phi^{-1}   + u  
 \end{equation} 
 \begin{equation}\label{RENCPII4} 
  \Phi^{'} \Phi^{-1} =  -2i \lambda   + u \chi \Phi^{-1}
 .\end{equation}
Now let consider the following substitution 

 \begin{equation}\label{RENCPII5} 
 \Gamma = \chi \Phi^{-1}
 \end{equation}
and after taking the derivation of above equation with respect to $z$ we get
\[\Gamma^{'} = \chi^{'} \Phi^{-1} - \chi \Phi^{-1} \Phi^{'} \Phi^{-1}.\]
 Finally by making use of $ \Gamma$  and $ \Gamma^{'} $ in  ( \ref{RENCPII3}) and (\ref{RENCPII4}) and after simplification  we obtain the following expression
\begin{equation}\label{RENCPII8} 
 \Gamma^{'} = -4i\lambda \Gamma +u - \Gamma u \Gamma
\end{equation} 
the  above equation (\ref{RENCPII8}) is NC PII Riccati form in $ \Gamma $ where $u$ is the solution of  NC PII equation (\ref{NCPIIa}).    As $\Gamma $ has been expressed in terms of $\chi $ and $ \Phi$, the components of eigenvector of NC PII system (\ref{NCPIIb}).\\ 
\textbf{Remark1.1.}\\
We can easily  show that the NC PII Riccati form (\ref{RENCPII8}) can be satisfied by taking  the solutions of  $ u $ and $  \Gamma $ as follow
\[ u [1]  = -8i\lambda_{1} (1- e^{-8i\lambda_{1} z})^{-1} e^{-4i\lambda_{1} z}\]
\[  \Gamma = e^{4i\lambda_{1} z} \]
in that equation. The following proposition 1.2. involves the zero curvature representation of non-linear differential equations (\ref{TODA2}) and (\ref{TODA3})\\ 
\textbf{Proposition 1.2.}\\
The compatibility condition of  linear systems
\begin{equation}\label{TODA4}
\left\{
\begin{array}{lr}
 \Psi_{\lambda}=L  \Psi  \\
  \Psi_{z}=M \Psi 
\end{array}
\right.
\end{equation} 
With the Lax matrices 
\begin{equation}\label{TODALP}
\left\{
\begin{array}{lr}
L=  2 \lambda^{2} I - q^{'} i \sigma_{2} + (- q^{2}- 2 \phi \psi ) \sigma_{3} - 4  z \Sigma \\
M =  q  \sigma_{1} + \lambda I  
\end{array}
\right. 
\end{equation} 
yields  equation  (\ref{TODA2}) when $ q= \phi $ and for $ q = \psi $ the compatibility condition gives equation   (\ref{TODA3}). Here $\sigma_{1}$, $\sigma_{2}$, $\sigma_{3}$ are Pauli spin matrices, $I$ is identity matrix of order $2$ and $ \Sigma  = \begin{pmatrix}
0 & 0 \\
 0 & 1
\end{pmatrix} .$ \\ 
 \textbf{Proof:}\\
 We can easily evaluate the following values
 \begin{equation}\label{TODA5}
A_{z} -B_{\lambda} = \left(\begin{array}{cc} -2 (\phi \psi)^{'} - (q^{2})^{'} -1  & - q^{''}  \\   q^{''}  &   2 (\phi \psi)^{'} + (q^{2})^{'} - 5  \end{array}\right)
\end{equation}
 \begin{equation}\label{TODA6}
BA -AB = \left(\begin{array}{cc}  q q^{'} +  q^{'} q & \omega_{+} \\  \omega_{-}  & -  q q^{'} -  q^{'} q \end{array}\right)
\end{equation}
where $ \omega_{+} = 2 q \phi \psi  + 2 q^{3 } -  4 q z + 2 \phi \psi q $ and  $ \omega_{-}  =  - 2 q \phi \psi  - 2q^{3 } + 4 z q- 2 \phi \psi q .$
Finally from zero curvature condition we get
\begin{equation}\label{TODA7}
2 (\phi \psi)^{'}  + 2 (q^{2} )^{'}  + 1 =0 
\end{equation}
\begin{equation}\label{TODA8}
q^{''} = - 2 q \phi \psi - 2 q^{3} -2 \phi \psi q + 4 q z 
\end{equation}
\begin{equation}\label{TODA9}
2 (\phi \psi)^{'}  + 2 (q^{2} )^{'}  - 5 =0 
\end{equation}
\begin{equation}\label{TODA10}
q^{''} = - 2 q \phi \psi - 2 q^{3}  - 2 \phi \psi q  + 4  z q 
\end{equation}
Now adding (\ref{TODA7}) and (\ref{TODA9})  , we get
\begin{equation}\label{TODA11}
 (\phi \psi)^{'}  +  (q^{2} )^{'}  -1 =0 
\end{equation}
on integrating  above emuation with respect to $z $ we get
\begin{equation}
 \phi \psi +  q^{2}  - z=D
\end{equation}
where $ D$ is constant of integration, set  $ D= 0 $ in above equation  then 
\begin{equation}\label{TODA12}
 \phi \psi +  q^{2}   - z= 0
\end{equation}
Now after combining  equation (\ref{TODA8}) and equation (\ref{TODA10})  we obtain
 \begin{equation}\label{TODA14} 
 q^{''} = - 2 q \phi \psi - 2 q^{3} -2 \phi \psi q  + 2 q z +  2  z q
\end{equation} \\
For $ q= \phi $ above expression  (\ref{TODA14})  

 \begin{equation}\label{TODA144} 
\phi^{''} = - 2  \phi (\phi  \psi +  \phi^{2} - z) -2 \phi \psi  \phi  + 2 z \phi.
\end{equation}
Now after using equation  (\ref{TODA12}) for $q =\phi $  in equation (\ref{TODA144}) we obtain following expression
 \begin{equation} 
\phi^{''} =2 z \phi -2 \phi \psi  \phi.  
\end{equation}
When $ q= \psi  $ then the  (\ref{TODA14}) can be written as
 \begin{equation}\label{TODA145} 
 \psi^{''} = - 2 \psi \phi \psi - 2( \psi^{2} + \phi \psi - z ) \psi   + 2 \psi z 
\end{equation}
again using equation  (\ref{TODA12}) for $q =\psi $  in above  (\ref{TODA145}, we get

 \begin{equation}\label{TODA145} 
 \psi^{''} = 2 \psi z  - 2 \psi \phi \psi  
\end{equation}\\
In next proposition 1.3., we derive the explicit expressions of  Darboux transformations for $ \phi$ and  $ \psi$ with the help of linear systems given in (\ref{TODA4}).

\textbf{Proposition 1.3.}\\
For the column vector $ \Psi=
\begin{pmatrix}
X  \\
Y
\end{pmatrix}$  in linear systems  (\ref{TODA4}) with the standard transformations on its components $X$ and $Y$ 
\begin{equation}\label{d3}
X \rightarrow X[1]=\lambda Y-\lambda_{1}Y_{1}(\lambda_{1})X_{1}^{-1}(\lambda_{1})X
\end{equation}
\begin{equation}\label{d4}
Y \rightarrow Y[1]=\lambda X-\lambda_{1}X_{1}(\lambda_{1})Y_{1}^{-1}(\lambda_{1})Y
\end{equation}
we can construct the Darboux transformations for $ \phi $  and $ \psi  $ as follow
\begin{equation}\label{d5}
 \phi[1] =  Y_{1} X_{1}^{-1} \phi Y_{1} X_{1}^{-1} 
\end{equation}
and 

\begin{equation}\label{d15}
 \psi[1] =  Y_{1} X_{1}^{-1} \psi Y_{1} X_{1}^{-1} 
\end{equation} 
respectively, where $ X $ , $ Y $ are arbitrary solutions at  $\lambda$ and  $X_{1}(\lambda_{1})$ ,  $Y_{1}(\lambda_{1})$ are the particular solutions at $ \lambda=\lambda_{1}$.\\
\textbf{Proof:}\\
Let us write the  second expression of (\ref{TODA4}) in the form of 
\begin{equation}\label{d6}
 \begin{pmatrix}
 X  \\
Y
\end{pmatrix}_{z} =
\begin{pmatrix}
  \lambda  & q \\
q &  \lambda
\end{pmatrix}
\begin{pmatrix}
 X  \\
 Y
\end{pmatrix}.
\end{equation}
Now under the  transformations (\ref{d3}) and (\ref{d4}) above equation (\ref{d6}) becomes
\begin{equation}\label{d7}
 \begin{pmatrix}
 X[1]  \\
Y[1]
\end{pmatrix}_{z} =
\begin{pmatrix}
  \lambda  & q[1] \\
q[1] & \lambda
\end{pmatrix}
\begin{pmatrix}
 X[1]  \\
 Y[1]
\end{pmatrix}.
\end{equation}
From (\ref{d6}) and (\ref{d7}) we obtain the following systems of equations
\begin{equation}\label{d8}
\left\{
\begin{array}{lr}
 X^{'} = \lambda X + q Y\\
 Y^{'} = \lambda Y + q X  
\end{array}
\right.
 \end{equation} 
and
\begin{equation}\label{d9}
\left\{
\begin{array}{lr}
 X^{'}[1] = \lambda X[1] + q[1] Y[1]\\
 Y^{'}[1] = \lambda Y[1] + q[1] X[1]  
\end{array}
\right.
 \end{equation} 
Now  after substituting the transformed values  $X[1] $ and  $Y[1] $ from (\ref{d3}) and (\ref{d4}) in equation (\ref{d9}) and then using (\ref{d8}) in resulting equation, we get  one fold Darboux transformation for $q $. \\ 
\begin{equation}\label{GDT}
 q[1] =  Y_{1} X_{1}^{-1} q Y_{1} X_{1}^{-1}. 
\end{equation} 
It is obvious that by taking $ q= \phi $ in (\ref{GDT}) we obtain  (\ref{d5}) and for   $ q= \psi $  we get transformation (\ref{d15}) on $ \psi $.  In upcoming section after taking a brief review of quasideterminant, we will substitute $ \phi $ and $ \psi$ as seed solutions in  NC PII Darboux transformations \cite{MIRFAN}. Finally we generalize the Darboux transformations  (\ref{d5}) and  (\ref{d15}) to the $N$-th form.
\section{A Brief Introduction of Quasideterminants}
This section is devoted to a brief review of quasideterminants introduced by Gelfand
and Retakh  \cite{GelRet}. Quasideterminants are the replacement for the determinant for matrices with noncommutative entries and these determinants plays very important role to construct the multi-soliton solutions of NC integrable systems \cite{r25}, \cite{r26}, by applying the Darboux transformation. Quasideterminants are not just a noncommutative generalization of usual commuta-
tive determinants but rather related to inverse matrices, quasideterminants for the square matrices are defined as
if $A = a_{ij}$ be a $ n \times n $ matrix and $B = b_{ij}$ be the inverse matrix of A. Here all
matrix elements are supposed to belong to a NC ring with an associative
product.
Quasideterminants of $A$ are defined formally as the inverse of the elements of $B = A^{-1}$
\[ |A|_{ij}=b^{-1}_{ij} \] this expression under the limit $ \theta^{\mu \nu} \rightarrow 0 $ , means entries of $A$ are commuting, will reduce to \[ |A|_{ij}= (-1)^{i+j}\frac{detA}{detA^{ij}} \]
where $A^{ij}$ is the matrix obtained from $ A $ by eliminating  the $ i$-th row and the $ j$-th column.
We can write down more explicit form of quasideterminants. In order to see it, let us
recall the following formula for a square matrix

\begin{equation} A =
\begin{pmatrix}
 A & B \\
C & D
\end{pmatrix}^{-1}
=\begin{pmatrix}
A-BD^{-1}C)^{-1} & -A^{-1}B(D-CA^{-1}B)^{-1} \\
-(D-CA^{-1}B)^{-1} CA^{-1} & (D-CA^{-1}B)^{-1}
\end{pmatrix}
\end{equation}
where $A$ and $D$ are square matrices, and all inverses are supposed to exist. We note that
any matrix can be decomposed as a $
2\times 2$ matrix by block decomposition where the diagonal
parts are square matrices, and the above formula can be applied to the decomposed $2 \times 2$
matrix. So the explicit forms of quasideterminants are given iteratively by the following
formula

\[ \vert A \vert _{ij}=a_{ij}-\Sigma_{p\neq i , q\neq j}  a_{iq} \vert {A}^{ij} \vert^{-1} _{pq} a_{pj}  \]
the  number of quasideterminant of a given matrix will be equal to the numbers of its elements for example a matrix of order $3$ has nine quasideterminants. It is sometimes convenient to represent the quasi-determinant as follows
\begin{equation} \label{QDD} 
 \vert A \vert_{ij}=\begin{vmatrix}
a_{11} &  \cdots  &  a_{1j}          & \cdots  & a_{1n}\\
\vdots & \vdots   & \vdots           & \vdots  & \vdots\\
a_{i1} & \cdots   & \boxed{a_{ij}}    & \cdots  & a_{in}\\
\vdots & \vdots   & \vdots           & \vdots  & \vdots\\
a_{in} & \cdots   & a_{ni}           & \cdots  & a_{nn}
\end{vmatrix}. \end{equation}
Let us consider  examples of matrices with order $2$ and $3$, for $2\times2$ matrix

\[ A = \begin{pmatrix}
          a_{11} & a_{12} \\
          a_{21} & a_{22} \\
       \end{pmatrix} \]
now the quasideterminats of this matrix are given below
\[  \vert A \vert _{11}=\begin{vmatrix}
{\boxed{a_{11}}} & a_{12} \\
a_{21} & a_{22}
\end{vmatrix} = a_{11} - a_{12} a^{-1}_{22} a_{21} \]

 \[  \vert A \vert _{12}=\begin{vmatrix}
a_{11} &  {\boxed{a_{12}}}\\
a_{21} & a_{22}
\end{vmatrix} = a_{12} - a_{22} a^{-1}_{21} a_{12} \]

\[  \vert A \vert _{21}=\begin{vmatrix}
a_{11} & a_{12}\\
 {\boxed{a_{21}}} & a_{22}
\end{vmatrix} = a_{21} - a_{11} a^{-1}_{12} a_{22} \]

 \[  \vert A \vert _{22}=\begin{vmatrix}
a_{11} & a_{12}\\
a_{21}& {\boxed{a_{22}}}
\end{vmatrix} = a_{22} - a_{21} a^{-1}_{11} a_{12}. \]
 The number of quasideterminant of a given matrix will be equal to the numbers of its elements for example a matrix of order $3$ has nine quasideterminants.
Now we consider the example of $3\times3$ matrix, its first quasidetermints can be evaluated in the following way\\

$ \vert A \vert _{11}= \begin{vmatrix}
{\boxed{a_{11}}} & a_{12} & a_{13}\\
 a_{21} & a_{22} & a_{23}\\
 a_{31} & a_{32} & a_{33}
\end{vmatrix}= a_{11}-a_{12} M a_{21}-a_{13} M a_{21}-a_{12} M a_{31}-a_{13} M a_{31}$\\
where $ M =\begin{vmatrix}
{\boxed{a_{22}}} &  a_{23}\\
a_{32} & a_{33}
\end{vmatrix}^{-1} ,$ similarly we can evaluate the other eight quasideterminants of this matrix.\\  
\\
\section{ Quasideterminant representation of Darboux transformation}
\subsection{Darboux transformations of NC PII equation}
In the theory of integrable systems the applications of Darboux transformations (DTs) are quite interesting to construct the multi-soliton solutions of these systems.  These transformations consist the  particular solutions of corresponding linear systems of the  integrable equations and their seed (initial) solutions. For example the NC PII equation (\ref{NCPIIa}) possesses following 
$N$ fold DT  
\begin{equation}\label{NCPDTS} 
 u[N+1]=\Pi^{N}_{k=1} \Theta_{k}[k] u \Pi^{1}_{j=N}\Theta_{j}[j] \;\;\;\; \text{for} \;\; N \geq 0
\end{equation} 
with 
\[\Theta_{N}[N]=\Lambda_{N}^{\phi}[N] \Lambda_{N}^{\chi}[N]^{-1}\]
where $u[1]$ is seed solution and $  u[N+1]$ are the new solutions of NC PII equation  \cite{MIRFAN}. In above transformations (\ref{DT1}) $ \Lambda_{N}^{\phi}[N] $ and $ \Lambda_{N}^{\chi}[N] $ are the quasideterminants of the particular solutions of NC PII linear system (\ref{NCPIIb}). Here the odd order quasideterminant representationS of $ \Lambda_{2N+1}^{\phi}[2N+1] $ and $ \Lambda_{2N+1}^{\chi}[2N+1] $  are presented below
 \[\Lambda_{2N+1}^{\phi}[2N+1]=
\begin{vmatrix}
 \Phi_{2N} & \Phi_{2N-1} & \cdots& \Phi_{1} & \Phi\\
  \lambda_{2N} \chi_{2N} &  \lambda_{2N-1} \chi_{2N-1} & \cdots &  \lambda_{1}\chi_{1} &  \lambda \chi\\
\vdots & \vdots & \cdots & \vdots & \vdots\\
  \lambda^{2N-1}_{2N} \chi_{2N} &  \lambda^{2N-1}_{2N-1} \chi_{2N-1} & \cdots &  \lambda^{2N-1}_{1}\chi_{1} &  \lambda^{2N-1} \chi\\
  \lambda^{2N}_{2N} \Phi_{2N} &  \lambda^{2N}_{2N-1} \Phi_{2N-1} & \cdots &  \lambda^{2N}_{1} \Phi_{1} & {\boxed{ \lambda^{2N} \Phi}}
\end{vmatrix} \]
and

 \[\Lambda_{2N+1}^{\chi}[2N+1]=
\begin{vmatrix}
 \chi_{2N} & \chi_{2N-1} & \cdots & \chi_{1} & \chi\\
 \lambda_{2N} \Phi_{2N} &  \lambda_{2N-1} \Phi_{2N-1} & \cdots &  \lambda_{1} \Phi_{1}&  \lambda \Phi\\
 \vdots & \vdots & \cdots & \vdots & \vdots\\
  \lambda^{2N-1}_{2N} \Phi_{2N} &  \lambda^{2N-1}_{2N-1} \Phi_{2N-1} & \cdots &  \lambda^{2N-1}_{1}\Phi_{1} &  \lambda^{2N-1} \Phi\\
  \lambda^{2N}_{2N} \chi_{2N} &  \lambda^{2N}_{2N-1} \chi_{2N-1} & \cdots &  \lambda^{2N}_{1} \chi_{1} & {\boxed{ \lambda^{2N} \chi}}
\end{vmatrix} \]
with  \[\Lambda_{1}^{\phi}[1]=\Phi_{1}, \; \;  \; \Lambda_{1}^{\chi}[1] =\chi_{1} \]
where $ \chi_{1}, \chi_{2}, \chi_{3}, ..., \chi_{N} $ and $ \Phi_{1}, \Phi_{2}, \Phi_{3}, ..., \Phi_{N}  $ are the solutions of system (\ref{RENCPII2}) at $ \lambda_{1}, \lambda_{2}, \lambda_{3},..., \lambda_{N}$.

\textbf{Proposition 1.4.}\\
By taking $ u =  u_{1} = \phi^{'} \phi^{-1}$ as a seed solution in (\ref{NCPDTS}) the  $ \phi$ solitonic solutions $ u[N]$ of NC PII equation can be expressed in terms of $ N$-fold quasideterminant Darboux transformation of $ \phi $as follow
\begin{equation}\label{DT1} 
 u[N+1]=\Pi^{N}_{k=1} \Theta_{k}[k]  \phi^{'}[N] \phi^{-1}[N] \Pi^{1}_{j=N}\Theta_{j}[j] \;\;\;\; \text{for} \;\; N \geq 1
\end{equation} 
and $ \phi[N] $ is given by 
\[\phi[N]=\Pi^{N-1}_{k=0} \varTheta_{N-k}[N-k] \phi \Pi^{0}_{j=N-1}\varTheta_{N-j}[N-j]\]
where $ \varTheta_{N}[N]=\varOmega_{N}^{Y}[N] \varOmega_{N}^{X}[N]^{-1}$ and $ \varOmega_{N}^{Y}[N] $, $\varOmega_{N}^{X}[N] $ represent quasideterminants of order $N$.\\
\textbf{Proof:}\\
The one fold NC PII Darboux transformation with seed solution $ u_{1} = \phi^{'} \phi^{-1}$ can be written as
\begin{equation}\label{d11}
u[1]=\Phi_{1}\chi_{1}^{-1} \phi^{'} \phi^{-1}  \Phi_{1}\chi^{-1}_{1}
\end{equation} 
where $ \Phi$ and $ \chi$ are the eigenvector components of linear systems associated to NC PII equation \cite{MIRFAN}. 
Now we can express the two fold Darboux transformation as follow
\begin{equation}\label{TFI}
u[2]=\Phi_{1} [1]\chi_{1}^{-1}[1] \phi^{'}[1] \phi^{-1}[1]  \Phi_{1} [1] \chi^{-1}_{1} [1].
\end{equation} 
here $ \phi[1] $ is given in equation (\ref{d5}). Now we consider the third solitonic solution of NC PII equation as under

\begin{equation}\label{TF2}
u[3]=\Phi_{1} [2]\chi_{1}^{-1}[2] \phi^{'}[2] \phi^{-1}[2]  \Phi_{1} [2] \chi^{-1}_{1} [2].
\end{equation} 
where \[  \phi[2]=Y[1]X^{-1}[1]  \phi [1]Y[1]X^{-1}[1].\]
In order express $\phi[2] $ in terms of quasideterminant, fisrt we write the  transformations (\ref{d3}) and (\ref{d4}) by using the definition (\ref{QDD})as under
\begin{equation}\label{qd1}
X [1]=
\begin{vmatrix}
 X_{1} & X_{0}\\
\lambda_{1} Y_{1} & {\boxed{\lambda_{0} Y_{0}}}
\end{vmatrix}
=\delta_{X}^{e}[1]
\end{equation}
and
\begin{equation}\label{qd2}
 Y [1]=
\begin{vmatrix}
 Y_{1} & Y_{0}\\
\lambda_{1} X_{1} & {\boxed{\lambda_{0} X_{0}}}
\end{vmatrix}=\delta_{Y}^{e}[1]
\end{equation}
We have taken $\lambda=\lambda_{0}$, $X=X_{0}$ and $Y=Y_{0}$  in order to generalize the transformations in $N$th form.
Further,  we can represent the transformations $X [2]$ and $Y [2]$ by quasideterminants
 \[X [2]=
\begin{vmatrix}
 X_{2} & X_{1} & X_{0}\\
\lambda_{2} Y_{2}&\lambda_{1} Y_{1} &\lambda_{0} Y_{0}\\
\lambda^{2}_{2} X_{2} &\lambda^{2}_{1} X_{1}& {\boxed{\lambda^{2}_{0} X_{0}}}
\end{vmatrix}=\varUpsilon_{X}^{o}[2] \] and
 \[Y [2]=
\begin{vmatrix}
 Y_{2} & Y_{1} & Y_{0}\\
\lambda_{2} X_{2} &\lambda_{1} X_{1} &\lambda_{0} X_{0}\\
\lambda^{2}_{2} Y_{2} &\lambda^{2}_{1} Y_{1} & {\boxed{\lambda^{2}_{0} Y_{0}}}
\end{vmatrix}=\varUpsilon_{Y}^{o}[2].\] here superscripts $e$ and $o$  of $\varUpsilon$ represent the even and odd order quasideterminants.  The $N$th transformations for $\varUpsilon_{X}^{o}[N]$ and $\varUpsilon_{Y}^{o}[N]$ in terms of  quasideterminants are given below
  \[\varUpsilon_{X}^{o}[N]=
\begin{vmatrix}
 X_{N} & X_{N-1} & \cdots & X_{1} & X_{0}\\
\lambda_{N} Y_{N} &\lambda_{N-1} Y_{N-1} & \cdots &\lambda_{1}X_{1}&\lambda_{0} Y_{0}\\
 \vdots & \vdots & \cdots & \vdots & \vdots\\
\lambda^{N-1}_{N} Y_{N} &\lambda^{N-1}_{N-1} Y_{N-1} & \cdots &\lambda^{N-1}_{1}X_{1} &\lambda^{N-1}_{0} Y_{0}\\
\lambda^{N}_{N} X_{N} &\lambda^{N}_{N-1} X_{N-1} & \cdots &\lambda^{N}_{1} X_{1} & {\boxed{\lambda^{N}_{0} X_{0}}}
\end{vmatrix}\]and
  \[\varUpsilon_{Y}^{o}[N]=
\begin{vmatrix}
 Y_{N} & Y_{N-1} & \cdots& Y_{1} & Y_{0}\\
\lambda_{N} X_{N} &\lambda_{N-1} X_{N-1} & \cdots &\lambda_{1}X_{1} &\lambda_{0} X_{0}\\
\vdots & \vdots & \cdots & \vdots & \vdots\\
\lambda^{N-1}_{N} X_{N} &\lambda^{N-1}_{N-1} X_{N-1} & \cdots &\lambda^{N-1}_{1}X_{1} &\lambda^{N-1}_{0} X_{0}\\
\lambda^{N}_{N} Y_{N} &\lambda^{N}_{N-1} Y_{N-1} & \cdots &\lambda^{N}_{1} Y_{1} & {\boxed{\lambda^{N}_{0} Y_{0}}}
\end{vmatrix} \]
here $N$ is to be taken as even.
in the same way we can write $N$th quasideterminant representations of $\varUpsilon_{X}^{e}[N]$ and $\varUpsilon_{Y}^{e}[N]$.
 \[\varUpsilon_{X}^{e}[N]=
\begin{vmatrix}
 X_{N} & X_{N-1} & \cdots & X_{1} & X_{0}\\
\lambda_{N} Y_{N} &\lambda_{N-1} Y_{N-1} & \cdots&\lambda_{1}X_{1} &\lambda_{0} Y_{0}\\
 \vdots & \vdots& \cdots & \vdots & \vdots\\
\lambda^{N-1}_{N} X_{N} &\lambda^{N-1}_{N-1} X_{N-1} & \cdots&\lambda^{N-1}_{1}X_{1} &\lambda^{N-1}_{0} X_{0}\\
\lambda^{N}_{N} Y_{N} &\lambda^{N}_{N-1} Y_{N-1} & \cdots &\lambda^{N}_{1} Y_{1} & {\boxed{\gamma^{N}_{0} Y_{0}}}
\end{vmatrix}\]and
\[\varUpsilon_{Y}^{e}[N]=
\begin{vmatrix}
 Y_{N} & Y_{N-1} & \cdots & Y_{1} & Y_{0}\\
\lambda_{N} X_{N} &\lambda_{N-1} X_{N-1} & \cdots &\lambda_{1}X_{1} &\lambda_{0} X_{0}\\
 \vdots & \vdots & \cdots & \vdots & \vdots\\
\lambda^{N-1}_{N} Y_{N}&\lambda^{N-1}_{N-1} Y_{N-1} & \cdots&\lambda^{N-1}_{1} Y_{1} &\lambda^{N-1}_{0} Y_{0}\\
\lambda^{N}_{N} X_{N}&\lambda^{N}_{N-1} X_{N-1}& \cdots &\lambda^{N}_{1} X_{1} & {\boxed{\gamma^{N}_{0} X_{0}}}
\end{vmatrix}.\]

Similarly, we can derive the expression for $N$-fold Darboux transformations for $ \phi $ by applying the  transformation (\ref{d5})  iteratively, for this pupose let us consider
\[  \phi[1]=\varOmega_{1}^{Y}[1]\varOmega_{1}^{X}[1]^{-1} \phi  \varOmega_{1}^{Y}[1]\varOmega_{1}^{X}[1]^{-1}\]
where  \[\varOmega_{1}^{Y}[1]=Y_{1} \]
\[\varOmega_{1}^{X}[1] =X_{1}\]
this is one fold Darboux transformation. The two fold Darboux transformation is given by
 \begin{equation}\label{it2} 
  \phi[2]=Y[1]X^{-1}[1]  \phi [1]Y[1]X^{-1}[1].
 \end{equation} 
 We may rewrite the equation (\ref{qd1}) and equation (\ref{qd2}) in the following forms
\[X [1]=
\begin{vmatrix}
 X_{1} & X_{0}\\
\lambda_{1} Y_{1} & {\boxed{\lambda_{0} Y_{0}}}
\end{vmatrix}
=\varOmega_{2}^{X}[2]
\]

 \[Y [1]=
\begin{vmatrix}
 Y_{1} & Y_{0}\\
\lambda_{1} X_{1} & {\boxed{\lambda_{0} X_{0}}}
\end{vmatrix}=\varOmega_{2}^{Y}[2].\]
and equation (\ref{it2}) may be written as
\[ \phi[2]=\varOmega_{2}^{Y}[2] \varOmega_{2}^{X}[2]^{-1}\varOmega_{1}^{Y}[1]\varOmega_{1}^{X}[1]^{-1} \phi \varOmega_{1}^{Y}[1]\varOmega_{1}^{X}[1]^{-1}\varOmega_{2}^{Y}[2] \varOmega_{2}^{X}[2]^{-1}\]
We can show that the  fourth solitonic solution $ u[4]$ will take the following form
\begin{equation}
u[4]=\Phi_{1} [3] \chi_{1}^{-1} [3]\phi^{'} [3] \phi^{-1} [3]  \Phi_{1} [3] \chi^{-1}_{1} []
\end{equation} 
where the three fold Darboux transform $ \phi[3]$ can be expressed as
\[\phi[3]=\varOmega_{3}^{Y}[3] \varOmega_{3}^{X}[3]^{-1}\varOmega_{2}^{Y}[2] \varOmega_{2}^{X}[2]^{-1}\varOmega_{1}^{Y}[1]\varOmega_{1}^{X}[1]^{-1} \phi \varOmega_{1}^{Y}[1]\varOmega_{1}^{X}[1]^{-1}\varOmega_{2}^{Y}[2] \varOmega_{2}^{X}[2]^{-1}\varOmega_{3}^{Y}[3] \varOmega_{3}^{X}[3]^{-1}.\]
here
 \[X [2]=
\begin{vmatrix}
 X_{2} & X_{1} & X_{0}\\
 \lambda_{2} Y_{2}& \lambda_{1} Y_{1} & \lambda_{0} Y_{0}\\
 \lambda^{2}_{2} X_{2} & \lambda^{2}_{1} X_{1}& {\boxed{\lambda^{2}_{0} X_{0}}}
\end{vmatrix}= \varOmega_{3}^{X}[3]\] and
 \[Y [2]=
\begin{vmatrix}
 Y_{2} & Y_{1} & Y_{0}\\
 \lambda_{2} X_{2} & \lambda_{1} X_{1} & \lambda_{0} X_{0}\\
 \lambda^{2}_{2} Y_{2} & \lambda^{2}_{1} Y_{1} & {\boxed{\lambda^{2}_{0} Y_{0}}}
\end{vmatrix}=\varOmega_{3}^{Y}[3].\] 

Finaly, by applying the  transformtion iteratively we can construct the $N$-th solitonic solution of NC PII $ (z,  \beta +n-1 ) $  equation in the following form
\begin{equation}
 u[N+1]=\Pi^{N}_{k=1} \Theta_{k}[k] \phi^{'}[N]  \phi^{-1}[N] \Pi^{1}_{j=N}\Theta_{j}[j] \;\;\;\; \text{for} \;\; N \geq 0
\end{equation} 
the $ N$ fold Darboux transformation $ \phi[N]$ given by
\[ \phi[N]=\varTheta_{N}[N]\varTheta_{N-1}[N-1]...\varTheta_{2}[2]\varTheta_{1}[1] \phi \varTheta_{1}[1]\varTheta_{2}[2]...\varTheta_{N-1}[N-1]\varTheta_{N}[N]\]
or

\[\phi[N]=\Pi^{N-1}_{k=0} \varTheta_{N-k}[N-k] \phi \Pi^{0}_{j=N-1}\varTheta_{N-j}[N-j]\]
where 

\[\varTheta_{N}[N]=\varOmega_{N}^{Y}[N] \varOmega_{N}^{X}[N]^{-1}.\]
Here we present only the  $N$th expression for odd order quasideterminants  $\varOmega_{2N+1}^{Y}[2N+1]$ and  $ \varOmega_{2N+1}^{X}[2N+1] $ 
 \[\varOmega_{2N+1}^{Y}[2N+1]=
\begin{vmatrix}
 Y_{2N} & Y_{2N-1} & \cdots& Y_{1} & Y_{0}\\
 \lambda_{2N} X_{2N} & \lambda_{2N-1} X_{2N-1} & \cdots & \lambda_{1}X_{1} & \lambda_{0} X_{0}\\
\vdots & \vdots & \cdots & \vdots & \vdots\\
 \lambda^{2N-1}_{2N} X_{2N} & \lambda^{2N-1}_{2N-1} X_{2N-1} & \cdots & \lambda^{2N-1}_{1}X_{1} & \lambda^{2N-1}_{0} X_{0}\\
 \lambda^{2N}_{2N} Y_{2N} & \lambda^{2N}_{2N-1} Y_{2N-1} & \cdots & \lambda^{2N}_{1} Y_{1} & {\boxed{\lambda^{2N}_{0} Y_{0}}}
\end{vmatrix} \]
and

 \[\varOmega_{2N+1}^{X}[2N+1]=
\begin{vmatrix}
 X_{2N} & X_{2N-1} & \cdots & X_{1} & X_{0}\\
\lambda_{2N} Y_{2N} &\lambda_{2N-1} Y_{2N-1} & \cdots & \lambda_{1}X_{1}& \lambda_{0} Y_{0}\\
 \vdots & \vdots & \cdots & \vdots & \vdots\\
 \lambda^{2N-1}_{2N} Y_{2N} & \lambda^{2N-1}_{2N-1} Y_{2N-1} & \cdots & \lambda^{2N-1}_{1}X_{1} & \lambda^{2N-1}_{0} Y_{0}\\
 \lambda^{2N}_{2N} X_{2N} & \lambda^{2N}_{2N-1} X_{2N-1} & \cdots & \lambda^{2N}_{1} X_{1} & {\boxed{\lambda^{2N}_{0} X_{0}}}
\end{vmatrix}.\]
Similarly, we can derive an explicit expression of $N$-fold Darboux transformation for $ \psi$ in the following form
\[\psi[N]=\Pi^{N-1}_{k=0} K_{N-k}[N-k] \psi \Pi^{0}_{j=N-1} K_{N-j}[N-j]\]
where $ K_{N}[N]=\varXi_{N}^{Y}[N] \varXi_{N}^{X}[N]^{-1}$ and $ \varXi_{N}^{Y}[N] $, $\varXi_{N}^{X}[N] $ represent quasideterminants of order $N$.
Similarly, we can construct the $N$-th soliton solution of NC PII$(z, \beta -n)$ equation in terms of quasideterminant by taking  $ u = u_{-1}  = \psi^{\prime} \psi^{-1} $ as a seed solution in its Darboux transformation (\ref{DT1}).  
In next sections we some basic quantum commutation relations  \cite{NH}, we will observe that in section 6 how these commutation relations  are helpful to derive  quantum PII equation from its Lax representation which involves  Planck constant $ \hbar  $ explicitly.      
\section{Quantum Painlev\'e II equation}
The  quantum extension of classical  Painlev\'e equations
involves the symmetrical form of Painlev\'e equations proposed in  \cite{NH} noncommuting objects. For the  quantum Painlev\'e II equation let us replace the function $u_{0}$ , $u_{1}$, $u_{2}$ by  $f_{0}$ , $f_{1}$, $f_{2}$ respectively in system  (\ref{eq:1}) , further parameters $  \alpha_{0} $ and $  \alpha_{1}$ belong to
the complex number field $\mathbb{C}$. The operators $f_{0}$, $f_{1}$ and $f_{2}$  obey the following commutation rules
\begin{equation}\label{qp1} 
[f_{1},f_{0}]_{-}= 2\hbar f_{2}, \; \; [f_{0},f_{2}]_{-}=[f_{2},f_{1}]_{-}= \hbar
\end{equation}
where $ \hbar  $ is Planck constant, the derivation $ \partial_{z}$ preserves the commutation relations (\ref{qp1}) \cite{NH} .
The NC differential system (\ref{eq:1})   admits the affine Weyl group
actions of type $\mathcal{A}_{l}^{(1)} $ and quantum PII equation 
\begin{equation}\label{qpII} 
 f^{''}_{2}=2f^{3}_{2}-zf_{2} + \alpha_{1}-\alpha_{0}. 
\end{equation}  
 can be obtained by elimination $f_{0}$ and $f_{1}$ from system  (\ref{eq:1})   with the help of  commutation relations (\ref{qp1}). The above equation (\ref{qpII}) is called quantum PII equation because 
 after eliminating  $f_{0}$ and $f_{2} $ from same system ( \ref{eq:1})   we obtain $P_{34}$  that involves Planck constant $\hbar $ \cite{NH}  and \cite{NGR}. In next section I construct a linear systems whose compatibility condition yields quantum PII equation with quantum commutation relation between function $ f_{2}$ and independent variable $ z$, further we show that under the classical limit when $ \hbar \rightarrow 0 $ this 
 system will reduce to classical PII equation.
\section{Zero curvature representation of quantum PII equation}
\textbf{Proposition 1.5.}\\
The compatibility condition of following linear system
\begin{equation}\label{QPIIb} 
 \Psi_{\lambda}=A(z;\lambda)\Psi , \; \; \;  \Psi_{z}=B(z;\lambda)\Psi 
\end{equation} 
with Lax matrices
\begin{equation}\label{RDTa}
\left\{
\begin{array}{lr}
A= (8i \lambda^{2} + if^{2}_{2} - 2iz ) \sigma_{3} + f^{'}_{2} \sigma_{2}+  (\frac{1}{4} c \lambda^{-1} -4\lambda f_{2} )\sigma_{1} + i \hbar \sigma_{2} \\
B =  -2i \lambda \sigma_{3}  + f_{2} \sigma_{1}  + f_{2}I
\end{array}
\right.
 \end{equation} 
yields quantum PII equation, here $I$ is $ 2 \times 2$ identity matrix and $ \lambda $ is spectral parameter and $c$ is constant.\\
\textbf{Proof:}\\
The compatibility condition of system (\ref{QPIIb}) yields zero curvature condition
\begin{equation}\label{ZC4} 
 A_{z}-B_{\lambda}= [B,A]_{-}.
\end{equation} 
 We can easily  evaluate the values for  $ A_{z}$, $B_{\lambda}$ and $[B,A]_{-} = BA -AB$ from the linear system (\ref{RDTa}) as follow
\begin{equation}\label{V1} 
 A_{z} = ( i f^{'}_{2} f_{2} + if_{2} f^{'}_{2}  -2i )\sigma_{3} + f^{''}_{2} \sigma_{2} - 4 \lambda f^{'}_{2} \sigma_{1} 
\end{equation} 

\begin{equation}\label{V2} 
 B_{\lambda} =  -2i \sigma_{3}  
\end{equation} 
and

\begin{equation}\label{V3} 
[ B,A ]=   \begin{pmatrix}
 i f^{'}_{2} f_{2} + if_{2} f^{'}_{2} + [f_{2},z]_{-} - i  \hbar  &  \delta  \\
 \lambda & - i f^{'}_{2} f_{2} - if_{2} f^{'}_{2} [z, f_{2}]_{-} + i  \hbar 
\end{pmatrix}
\end{equation} 

where
\[ \delta =  -i f^{''}_{2}+2if_{2}^{3}-2i[z,f_{2}]_{+} +ic  + i [ f^{'}_{2}, f_{2}]_{-} + 4i\lambda  \hbar  \]

and

\[  \lambda = i f^{''}_{2}- 2if_{2}^{3}+2i[z,f_{2}]_{+} - ic  + i [f_{2}, f^{'}_{2} ]_{-}- 4i\lambda   \hbar.\]
now after substituting these values from (\ref{V1}), (\ref{V2}) and (\ref{V3}) in equation  (\ref{ZC4}) we get

\begin{equation}\label{RM1} 
 \begin{pmatrix}
[ f_{2},z]_{-} - i  \hbar  &  \delta  \\
 \lambda & [z, f_{2}]_{-} + i  \hbar 
\end{pmatrix}=0
\end{equation} 
and the above result (\ref{RM1}) yields the following expressions   
\begin{equation}\label{L4} 
 [ f_{2},z] = \frac{1}{2}i  \hbar f_{2} 
\end{equation} 
and 
\begin{equation}\label{L5} 
 i f^{''}_{2}- 2if_{2}^{3} + 2i[z,f_{2}]_{+} - ic  + i [f_{2}, f^{'}_{2} ]_{-} - 4i\lambda \hbar=0
\end{equation} 
equation (\ref{L4}) shows quantum relation between the variables $ z$ and $ f_{2}$. In equation (\ref{L5}) the term $ i [f_{2}, f^{'}_{2}]_{-} - 2i\lambda \hbar $ 
can be  eliminated  by using  equation $ f^{'}_{2} = f_{1} - f_{0}$ from  (\ref{eq:1} )   and  quantum commutation relations (\ref{qp1}). For this purpose let us replace 
$ f_{2}  $ by $ -  \frac{1}{2}\lambda^{-1}  f_{2}$ in  (\ref{qp1}),  then commutation relations become 
\begin{equation}\label{L6} 
[f_{0},f_{2}]_{-}=[f_{2},f_{1}]_{-}=  -2 \lambda \hbar.
\end{equation}
Now let us take the commutator of the both side of the equation  $ f^{'}_{2} = f_{1} - f_{0}$  with $ f_{2}$ from right  side, we get

\[ [ f^{'}_{2}, f_{2} ]_{-} = [ f_{1}, f_{2}]_{-} - [ f_{0} , f_{2}]_{-} \]
above equation  with the commutation relations (\ref{L6}) can be written as
\begin{equation}\label{L7} 
  [ f^{'}_{2}, f_{2} ]_{-}= - 4 \lambda \hbar .
\end{equation}
Now after substituting the value of $   [ f^{'}_{2}, f_{2} ]_{-} $ from (\ref{L7}) in (\ref{L5}) we get 
\[  i f^{''}_{2}- 2if_{2}^{3} + 2i[z,f_{2}]_{+} - ic =0.\]
Finally, we can say that  the compatibility of condition of linear system (\ref{QPIIb}) yields the following expressions
\begin{equation}\label{L8} 
\left\{
\begin{array}{lr}
 f^{''}_{2} = 2 f^{3}_{2}- 2[z,f_{2}]_{+} + c \\
 z f_{2} - f_{2} z =  i \hbar f_{2}
\end{array}
\right.
 \end{equation} 
in above system (\ref{L8}) the first equation can be considered as a pure version of quantum Painlev\'e II equation that is equipped with  a quantum commutation relation $[ z, f_{2}]_{-} = - i  \hbar$  and this 
equation can be  reduced to the classical PII equation under the classical  limit when $  \hbar \rightarrow 0$.\\

\textbf{Remark 1.2.}\\
The linear system (\ref{QPIIb}) with eigenvector $ \Psi = \begin{pmatrix}
\psi_{1}  \\
\psi_{2} 
\end{pmatrix}$ and setting $ \varDelta = \psi_{1} \psi^{-1}_{2}$ can be reduced to the following quantum PII Riccati form
  \[  \varDelta^{'} = -4i \varDelta + f_{2}  + [ f_{2},\varDelta ]_{-}  - \varDelta f_{2}  \varDelta \]
\textbf{Proof:}\\
Here we apply the method of Konno and Wadati \cite{KKMW} to the linear  system (\ref{L8}) of quantum PII equation. For this purpose let us substitute the eigenvector $ \Psi = \begin{pmatrix}
\psi_{1}  \\
\psi_{2} 
\end{pmatrix}$ in linear systems (\ref{QPIIb}) and  we get
\begin{equation}\label{NCQPIIa}  
 \left \{ \begin{aligned}
\frac{d\psi_{1}} {d\lambda} = ( 8  i \lambda^{2} +  i f^{2}_{2}-2i z) \psi_{1} + ( - i f^{'}_{2}+\frac{1}{4} C_{0} \lambda^{-1}-4 \lambda f_{2}  + \hbar ) \psi_{2}\\
\frac{d\psi_{2}}{d\lambda} = (i f^{'}_{2} +\frac{1}{4} C_{0} \lambda^{-1}-4\lambda f_{2} - \hbar) \psi_{1} + (-8  i\lambda^{2} - i f^{2}_{2} +2i z)\psi_{2}
\end{aligned}
 \right. 
 \end{equation}
and 
\begin{equation}\label{NCQPIIb}  
\left \{ \begin{aligned}
\psi_{1}^{'} =( -2i \lambda + f_{2} ) \psi_{1} + f_{2} \psi_{2}\\
\psi_{2}^{'} =f_{2} \psi_{1} + ( 2i \lambda  + f_{2} )\psi_{2}
\end{aligned}
 \right. 
\end{equation}
where $ \psi_{1}^{'} = \frac{d \psi_{1}}{dz}$ and  now from system (\ref{NCQPIIb}) we can derive the following expressions 
 \begin{equation}\label{NCQPIIc} 
  \psi_{1}^{'} \psi_{2}^{-1} =  (-2i \lambda  + f_{2}  ) \psi_{1} \psi^{-1}_{2}   +  f_{2}   
 \end{equation} 
 \begin{equation}\label{NCQPIId} 
  \psi_{2}^{'} \psi_{2}^{-1} =  -2i \lambda  + f_{2}  + f_{2}  \psi_{1} \psi^{-1}_{2}.
 \end{equation}
Let consider the following substitution 
\begin{equation}\label{NCQPIIe} 
 \varDelta = \psi_{1} \psi^{-1}_{2}
 \end{equation}
now taking the derivation of above equation with respect to $z$
\[\varDelta^{'} = \psi^{'}_{1} \psi^{-1}_{2} - \psi_{1} \psi^{-1}_{2} \psi^{'}_{2} \psi^{-1}_{2}\]
after using the ( \ref{NCQPIIc}), (\ref{NCQPIId}) and (\ref{NCQPIIe}) in above equation we obtain
\begin{equation}\label{NCQPIIf} 
 \varDelta^{'} = -4i \varDelta + f_{2}  + [ f_{2},\varDelta ]_{-}  - \varDelta f_{2}  \varDelta
\end{equation} 
the  above expression (\ref{NCQPIIf}) can be considered as quantum Riccati equation in $ \varDelta$ because it involves commutation $[ f_{2},\varDelta ]_{-} = f_{2} \varDelta - \varDelta f_{2}  $  that has been derived from the linear system (\ref{NCQPIIb}).\\
\\

\section{Conclusion}
In this paper, I have derived non-vacuum solutions of NC PII equation taking the solutions of Toda equations at $ n=1$ as seed solutions in its Darboux transformation. I have also generalized the Darboux transformations of these seed solutions to $N$-th form.
Further, I derived a zero curvature representation  of quantum Painlev\'e II equation with its associated Riccati form and  also we  have derived an explicit expression of NC PII Riccati equation from the linear system of NC PII equation by using the method of Konno and Wadati  [\cite{KKMW}].  Further, one can derive B\"cklund transformations for NC PII equation with the help of our NC PII linear system and  its Riccati form by using the technique described in \cite{KKMW, AROD}, these transformations may be helpful to construct the nonlinear principle of superposition for NC PII solutions. It also  seems interesting to construct the connection of NC PII equation to the known integrable systems such as its connection to NC nonlinear Schr\"odinger equation and 
to NC KdV equation as it possesses this property in classical case. Further it is quite interesting symmetrically to construct zero curvature representations for quantum Painlev\'e equation PIV, PV in such a way to derive the similar results that have been described in (\ref{L8}) for quantum PII by using the quantum commutation relations of \cite{NH}.  
\section{Acknowledgement}
I would like to thank V. Roubtsov and V. Retakh  for their  discussions to me during my  Ph. D. research work.
 My special thanks to the University of the Punjab, Pakistan, on funding me for my Ph.D. project in France. I am also thankful to LAREMA, Universit\'e d'Angers and ANR "DIADEMS", France on providing me facilities during my Ph.D. research work.


\end{document}